\documentclass{ws-procs11x85}
\usepackage{multicol}
\usepackage{amsfonts}
\usepackage{amsthm}
\usepackage{amssymb}
\usepackage[mathscr]{euscript}
\usepackage{t1enc}
\usepackage{amsmath}
\usepackage{amscd}
\usepackage{psfrag}
\usepackage{graphics}
\begin{document}
\title{Direct quantization of equations of motion:\\
${}$\\
from classical dynamics to transition amplitudes via strings}
\author{Denis Kochan\footnote{Author is looking for a postdoctoral position in the area of mathematical physics.}}
\address{Department of Theoretical Physics and Physics Education\\
FMFI UK, Mlynsk\' a dolina F2, 842 48 Bratislava, Slovakia\\
\vspace{0.3cm}
\texttt{kochan@fmph.uniba.sk}}
\date{}
\maketitle
\abstract{New method of quantization is presented. It is based on classical Newton-Lagrange equations of motion
(representing the fundamental physical law of mechanics) rather than on their traditional Lagrangian and/or Hamiltonian
precursors.
It is shown that classical dynamics is governed by canonical two-form $\Omega$, which embodies kinetic energy and forces
acting within the system. New type of variational principle employing differential two-form $\Omega$ and
``\,umbilical strings\,'' is introduced. The Feynman path integral over histories of the system is then rearranged to
``\,umbilical world-sheet\,'' functional integral in accordance with the proposed variational principle. In the case of
potential-generated forces, world-sheet approach reduces to the standard quantum mechanics. As an example
\emph{Quantum Mechanics with friction} is analyzed in detail.\\

\noindent\textbf{PACS:} 01.70.+w, 02.40.Yy, 03.65.Ca, 45.20.-d\\

\noindent\textbf{Keywords}: \emph{quantization of dissipative systems, umbilical strings, path \emph{vs.} surface integral}\\

\centerline{{Dedicated to my father.}}}

\section{Introduction}

Description and detailed understanding of classical and quantal phenomena attract attention of many physicists and
mathematicians for a long time. Classical and quantum mechanics are the best elaborated, examined and
understood parts of physics. Their mathematical setting is concentrated around beautiful and powerful
artillery, which includes differential geometry, functional analyzes, spectral calculus, group and
representation theory, (co)homology techniques and so on.

The aim of the paper is to provide a simple geometrical picture of classical and quantum mechanics for physical systems,
where Lagrangian and/or Hamiltonian description is missing. The central object in our approach is a certain canonical
two-form $\Omega$, which is defined in an extended tangent bundle. Its main properties are narrowly studied in
sections 2 and 3. The differential two-form $\Omega$ serves as a guide for a new type of variational principle.
In section 4 we introduce the notion of ``\,umbilical world-sheet.'' It generalizes the concept of the history of the
system and therefore it becomes important in the context of quantization. Variation uncovers desired classical trajectory
and, as a bonus, also some kind of minimal surface. In section 5 we will see how the ``\,umbilical strings\,'' can be used
to rearrange the Feynman integral over the histories of the system to the surface
functional integral. String formulation embodies a big advantage, it concerns
components of the forces rather than their potential. Main message of the ``\,umbilical approach\,'' can be summarized
as follows: \emph{dissipative time evolution remains unitary, but it does not form a group(oid) according to composition}.
In section 6 we succeed to compute transition probability
amplitude for quantum system with friction performing explicitly the surface functional integration. For the
potential-generated forces, the ``\,umbilical world-sheet\,'' approach reduces to the standard quantum mechanics.

This paper is hopefully more readable and elaborated version of the article.\cite{kochan} There, we focused on rigorous
geometrical formulation of the theory. In the actual paper, we prefer to explain the main idea and exhibit its
functionality.

To be honest and collegial, it is necessary to bring some standard references on the quantization
of dissipative systems.\cite{kanai}${}^-$\cite{dodonov} The reader can find there material that can be compared
with our results.

\section{Lagrangian mechanics and null-spaces of the distinguished two-form $\Omega$}

Physical content of classical mechanics is represented by the second Newton's law. Its mathematical formulation
coincides in general curvilinear coordinates with Lagrange equations:\cite{arnold}${}^-$\cite{fecko1}
\begin{equation}\label{Lagrange eq.}
\frac{d}{dt}\left(\frac{\partial\,\mathbb{T}}{\partial\, \dot{q}^{\,i}}\right)
-\frac{\partial\,\mathbb{T}}{\partial\, q^{i}}=\mathbb{Q}\,_i\ \ \ \ \
i=1,\dots, n(=\mbox{the number of degrees of freedom)}\,.
\end{equation}
Here, $\mathbb{T}(q,\dot{q},t)$ is the kinetic energy of the system and $\mathbb{Q}\,_{i}(q,\dot{q},t)$ is the $i$-th component of a generalized
force. In the special case, when forces are potential-generated
$$
\mathbb{Q}\,_{i}=-\frac{\partial\,\mathbb{U}}{\partial\, q^{i}}+\frac{d}{dt}\left(\frac{\partial\,\mathbb{U}}{\partial\, \dot{q}^{\,i}}\right)\,,
$$
one can introduce the Lagrangian function $\mathbb{L}=\mathbb{T}-\mathbb{U}$ and write down the celebrated Euler-Lagrange
equations.
Generalized coordinates $\{q^{i}\}$ cover some open patch of the configuration space ($n$-dimensional manifold)
$M$.

If we restrict ourselves to the Lagrangian picture, the space of all physical states is the set of all admissible initial
conditions for the differential system (\ref{Lagrange eq.}). Geometrically, initial condition specified at the time $t_0$ by
the generalized position $q(t_0)=q_0$ and velocity $\dot{q}(t_0)=v_0$, defines a point $(q_0, v_0, t_0)$ in an extended
tangent bundle $TM\times\mathbb{R}$. Here an open patch of the extended tangent bundle\footnote{To be rigorous: a correct
geometrical setting for mechanics in the Lagrangian picture is represented by the line element contact bundle
$\mathscr{C}(M\times\mathbb{R})$ of the extended configuration space $M\times\mathbb{R}$.
The extended tangent bundle is its open dense submanifold, concisely, point
$\{q^{1},\dots,q^{n},v^{1},\dots,v^{n},t\}$ $\Leftrightarrow$ \emph{contact point}
$q=\{q^{1},\dots,q^{n},t\}\in M\times\mathbb{R}$ + \emph{line element}
$\ell=\mathrm{span}\{v^i\partial_{q^i}\bigr|_{q}+1\partial_t\bigr|_{q}\}\subset T_q(M\times\mathbb{R})$.} is dressed by the
$(2n+1)$-tuple of local coordinates $\{q^{1},\dots,q^{n},v^{1},\dots,v^{n},t\}$.

Let us destroy the compact form of (\ref{Lagrange eq.}) expressing generalized accelerations as functions of
the remaining entries:
$$
\ddot{q}^{\,i}=f^i(q,\dot{q},t,\mathbb{Q})\equiv\biggl(\frac{\partial^2\,\mathbb{T}}{\partial\, \dot{q}^{\,i}\,\partial\, \dot{q}^{\,a}}\biggr)^{-1}
\Bigl\{
\mathbb{Q}\,_{a}(q,\dot{q},t)+\frac{\partial\,{\mathbb{T}}}{\partial\, q^{a}}-
\frac{\partial^2\,\mathbb{T}}{{\partial\, \dot{q}^{\,a}}\,\partial\, q^{b}}\,\dot{q}^{\,b}-
\frac{\partial^2\,\mathbb{T}}{\partial\, \dot{q}^{\,a}\,\partial\, t}
\Bigr\}\,.
$$
When identifying $\dot{q}^{\,i}$ with $v^i$, we get instead of (\ref{Lagrange eq.}) the system of $2n+1$ first-order
differential equations:
\begin{equation}\label{Lagrange eq.2}
\dot{q}^{\,i}=v^i\,,\ \ \ \ \ \ \ \ \dot{v}^i=f^i(q,v,t,\mathbb{Q})\,,\ \ \ \ \ \ \ \ \dot{t}=1\,.
\end{equation}
The system above can be interpreted as a coordinate expression of a vector field on the extended
tangent bundle $TM\times\mathbb{R}$; down-to-earth, according to (\ref{Lagrange eq.2}) one can assign to any physical
state $(q,v,t)$ a tangent vector
\begin{equation}\label{Lagrange fld.}
\Gamma\Bigr|_{(q,v,t)}=\partial_{t}\Bigr|_{(q,v,t)} + v^i\,\partial_{q^i}\Bigr|_{(q,v,t)} + f^i(q,v,t,\mathbb{Q})\,\partial_{v^{i}}\Bigr|_{(q,v,t)}\in\ T_{(q,v,t)}\bigl(TM\times\mathbb{R}\bigr)\,.
\end{equation}

\begin{figure}[tbh]
\begin{center}
\psfrag{TQ}{\footnotesize{$TM$}}
\psfrag{TQxR}{\footnotesize{$TM\times\mathbb{R}$}}
\psfrag{QxR}{\footnotesize{$M\times\mathbb{R}$}}
\psfrag{Om}{\footnotesize{$\Omega$}}
\psfrag{R}{\footnotesize{$\mathbb{R}$}}
\psfrag{Q}{\footnotesize{$M$}}
\psfrag{t0}{$t$}
\psfrag{v0}{$v$}
\psfrag{q0}{$q$}
\psfrag{l}{$\ell$}
\psfrag{g}{$\gamma$}
\psfrag{G}{\footnotesize{$\Gamma\,\Bigr|_{(q,v,t)}$}}
\epsfxsize=8cm
\epsfbox{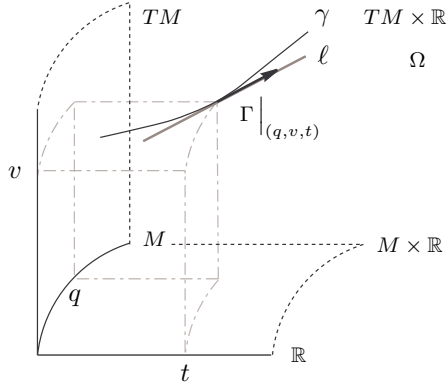}
\hspace{-3cm}
\caption{Time evolution in the Lagrangian picture is bedded in the extended tangent bundle. At each physical
state $(q,v,t)$ there is uniquely prescribed vector $\Gamma\,\bigr|_{(q,v,t)}$, which defines dynamics. Following its
integral curves, the complete time evolution is recovered.}\label{obraz}
\end{center}
\end{figure}

The time evolution with prescribed initial condition $(q_0, v_0, t_0)$ is represented by a curve in the extended tangent
bundle
$$
\gamma:\,\mathbb{R}[\tau]\rightarrow TM\times\mathbb{R}\,,\ \ \ \ \tau\mapsto\gamma(\tau)\equiv\bigl(q=q(\tau), v=v(\tau), t(\tau)=\tau\bigr)
$$
which passes through this physical state and, moreover, such that the tangent vector $\tfrac{d}{d\tau}\,\gamma(\tau)|_{\gamma(\tau)}$
at any point of the admissible trajectory equals to $\Gamma|_{\gamma(\tau)}$, see Figure \ref{obraz}.

We have just observed that classical dynamics is determined by the extended tangent bundle vector field $\Gamma$.
Having the function $\mathbb{T}(q,v,t)$ and the components of the generalized force $\mathbb{Q}(q,v,t)$ we can establish
(a priori by hand, but hopefully it will be soon clear that there are some very good reasons for that) the
two-form\footnote{Importance of the two-form $\Omega$ in context of mechanics is emphasized in the Arno\v{l}d's classical
monograph.\cite{arnold} However, I should mention the notice of Tam\' as F\" ul\" op. He drew my attention to the
book\cite{matolcsi} of Tam\'{a}s Matolcsi, where on the page 35 there is an equivalent expression for
the two-form.}

\begin{equation}\label{Omega}
\Omega:=\bigl\{\mathbb{Q}\,_i\,dq^i\bigr\}\wedge\,dt+d\Bigl\{\mathbb{T}\,dt+(\partial_{\,v^{i}}\,\mathbb{T})\,\bigl\{dq^i-v^i\,dt\bigr\}\Bigr\}\,.
\end{equation}
The main properties of $\Omega$ can be summarized as follows:
\begin{itemize}
\item[$\circ$] it is a differential two-form on the extended tangent bundle $TM\times\mathbb{R}$
\item[$\circ$] for any point $(q,v,t)$, it gives the linear map $\lrcorner:T_{(q,v,t)}\bigl(TM\times\mathbb{R}\bigr)\rightarrow T^*_{(q,v,t)}\bigl(TM\times\mathbb{R}\bigr)\,,
\mathrm{w}\mapsto\alpha:=\Omega(\mathrm{w},.)\equiv \mathrm{w}\lrcorner\,\Omega$
\item[$\circ$] if $\mathbb{T}$ is regular $\{\,\Leftrightarrow$ $\partial^2_{v^{i}v^{a}}\,(\mathbb{T})$ is invertible$\,\}$,
then the kernel $\ell$ of the contraction $\lrcorner$ is one-dimensional and it is spanned by the vector
$\Gamma|_{(q,v,t)}$; the subspace $\ell$ is called the null-space of the two-form $\Omega$
\item[$\circ$] whenever $\mathbb{Q}$ is potential-generated $\{\,\Leftrightarrow\,
\mathbb{Q}\,_i=-\tfrac{\partial\,\mathbb{U}}{\partial\, q^{i}}+\tfrac{d}{dt}\left(\tfrac{\partial\,\mathbb{U}}{\partial\,v^i}\right)\,\}$,
then $\Omega$ is exact, i.e. $\Omega=d\theta_{\,\mathbb{L}}$, where
$$
\theta_{\,\mathbb{L}}:=\mathbb{L}\,dt+(\partial_{\,v^{i}}\,\mathbb{L})\,\bigl\{dq^i-v^i\,dt\bigr\}
$$
is the well-known Lepage one-form on the extended tangent bundle that is associated to the Lagrangian function
$\mathbb{L=T-U}$
\item[$\circ$] performing Legendre transformation: $\bigl(q,v,t,\mathbb{L}\bigr)\mapsto \bigl(q,p=\tfrac{\partial\,{\mathbb{L}}}{\partial\,v},t,\mathbb{H}=p_iv^i-\mathbb{L}\bigr)$,
the one-form $\theta_{\,\mathbb{L}}$ transforms to the canonical Cartan-Poincar\`{e} one-form
$$
\omega_{\,\mathbb{H}}:=p_i\,dq^i-\mathbb{H}\,dt
$$
over the extended phase space $T^*M\times\mathbb{R}$
\item[$\circ$] the two-form $\Omega$, as well as the one-forms $\theta_{\,\mathbb{L}}$ and $\omega_{\,\mathbb{H}}$ are invariant
with respect to the group of diffeomorphisms\footnote{One easily verifies that everything remains invariant also
with respect to ``\,space-time\,'' transformations: $(q^i,t)\mapsto(Q^i(q,t),T=t)$.} of the configuration space $M$
\end{itemize}
So we could claim: \emph{Lagrangian mechanics is determined by the null-spaces of the distinguished two-form $\Omega$}.
Finding them, it is enough to pick up at each null subspace $\ell$ a vector $\mathrm{w}$ for which $\mathrm{w}\,\lrcorner\,dt=1$.
Doing this we are point-wisely reconstructing the dynamical vector field (\ref{Lagrange fld.}). Its integral curves are
solutions of the Lagrange equations.

\section{An outline: inverse problem of the calculus of variation. Quantization ambiguity}\label{section}

Before going to variational principle and quantization let us contemplate the inverse problem of the
calculus of variation. Roughly speaking the task is the following: assume that there is a system of the Newton-Lagrange
equations:
$$
\Bigl\{\dot{q}^{\,i}=v^i\ \ \mbox{and}\ \ \dot{v}^i=f^i(q,v,t)\Bigr\}\ \ \Longleftrightarrow\ \
\frac{d}{dt}\left(\frac{\partial\,\mathbb{T}}{\partial\, v^i}\right)
-\frac{\partial\,\mathbb{T}}{\partial\, q^{i}}=\mathbb{Q}\,_i\,,\ \ \mbox{where}\ \ \mathbb{T}=\frac{1}{2}v^i\,\delta_{ij}\,v^j
\ \ \mbox{and}\ \ \mathbb{Q}\,_i=f^i(q,v,t)\,.
$$
Does there exist a Lagrangian $\mathbb{L}$, whose variational equations $\tfrac{\delta\,\mathbb{L}}{\delta\,q}=0$ are
equivalent to the initial differential system? This problem was studied by many authors, for more
detail see the papers.\cite{henneaux1}${}^-$\cite{henneaux2} In what follows we will shortly outline that the answer
can be provided in terms of the distinguished two-form. Concisely, first of all we need to construct from
$\mathbb{T}=\tfrac{1}{2}v^i\,\delta_{ij}\,v^j$ and $\mathbb{Q}\,_i=f^i(q,v,t)$ the two-form $\Omega$. Since
the full dynamics is specified by its null-spaces, there is ``\,one degree of freedom\,'' in scaling:
$\Omega\mapsto\Omega^\prime=h(q,v,t)\,\Omega$, which does not affect them. Here $h(q,v,t)$ is some
(at least locally) non-zero function over the extended tangent bundle. If one succeeds to adjust the function $h$ in
such away that it will act as an integrator, $d\Omega^\prime=dh\,\wedge\,\Omega+h\,d\Omega=0$, then there should exist
a local $\Omega^\prime$-potential $\vartheta$ ($\Omega^\prime=d\vartheta$). It still remains to verify, whether there
exists a function $\mathbb{L}$ that satisfies
$\vartheta=\mathbb{L}\,dt+(\partial_{\,v^{i}}\,\mathbb{L})\,\bigl\{dq^i-v^i\,dt\bigr\}$. If yes, then the initial
differential system is ``\,derivable\,'' from that Lagrangian.

It is well-known since Darboux that in the one-dimensional ``\,world\,'' there always exists an appropriate Lagrangian.
As an example let us analyze a particle moving in $M=\mathbb{R}[x]$, driven by a force $\mathbb{Q}=\mathbb{Q}(x,v,t)$.
The integrator $h(x,v,t)$ of the two-form $\Omega=\mathbb{Q}\,dx\,\wedge\,dt+dv\,\wedge\,dx-v\,dv\,\wedge\,dt$,
for the system under the consideration, should obey the following partial differential equation:
\begin{equation}\label{integrator}
d\Omega^\prime=0\ \ \ \ \Longleftrightarrow\ \ \ \ \frac{\partial\,h}{\partial\,t}+v\,\frac{\partial\,h}{\partial\,x}+\mathbb{Q}\,\frac{\partial\,h}{\partial\,v}=-h\,\frac{\partial\,\mathbb{Q}}{\partial\,v}\,.
\end{equation}
To find all solutions of the above equation, one needs to provide characteristics (integral curves) of the
linear operator (vector field):
$
\partial_{t}+v\,\partial_{x}+\mathbb{Q}\,\partial_{v}-\{h\,\tfrac{\partial\,\mathbb{Q}}{\partial\,v}\}\,\partial_{h}
$
in $\mathbb{R}^4[x,v,t,h]$. Suppose that the characteristics are known:
$$
t=t(p)\,,\ \ \ x=x(p)\,,\ \ \ v=v(p)\,,\ \ \ h=h(p)\,,\ \ \ \mbox{where}\ p\ \mbox{is some flow parameter}\,.
$$
Then taking any function $z=z(x,v,t,h)$ such that $z(p)=z(x(p),v(p),t(p),h(p))$ is constant on these characteristics,
and solving the implicit function problem $z(x,v,t,h(x,v,t))=0$ with respect to $h$, we get all solutions of the initial
problem (\ref{integrator}).

For concreteness, let $\mathbb{Q}(x,v,t)=-v$ (a particle with friction that is linear in the actual velocity), then the
integral curves of the vector field
$\partial_{t}+v\,\partial_{x}-v\,\partial_{v}+h\,\partial_{h}$
are summarized as follows:
$$
t(p)=p+k_1\,,\ \ \ \ \ \ \ x(p)=-k_2\,\mathrm{e}^{-p}+k_3\,,\ \ \ \ \ \ \ v(p)=k_2\,\mathrm{e}^{-p}\,,\ \ \ \ \ \ \ h(p)=k_4\,\mathrm{e}^{\,p}\,,
$$
the numbers $(k_1,k_2,k_3,k_4)$ appear here as unimportant integral constants. For example for any real
numbers $\alpha$ and $\beta$ the function $z_{\alpha\beta}(v,t,h)=\mathrm{e}^{\alpha t}\,v^\beta\,h^{\beta-\alpha}-1$
is constant along any of the above characteristics. Let us, for simplicity, only consider the special case when
$\beta=\alpha-1$; then we can write
$$
\mbox{integrator:}\ \,h(v,t,\alpha)=\mathrm{e}^{\alpha\,t}v^{\alpha-1}\ \ \ \ \Longrightarrow\ \ \ \
\mbox{Lagrangian:}\ \begin{cases} \alpha\in\mathbb{R}/\{-1,0\}\,, & \mathbb{L}_{\alpha}=\tfrac{v^{\alpha+1}}{\alpha(\alpha+1)}\,\mathrm{e}^{\,\alpha\,t} \\
                                 \alpha=0\,, & \mathbb{L}_0=v\ln{v}-v-x \\
                                 \alpha=-1\,, & \mathbb{L}_{-1}= -\mathrm{e}^{-t}\ln{v}
                    \end{cases}
$$
Another type of solution is provided for example by the function $h=\tfrac{x}{v}+1$. So one can
see that for this relatively simple one-dimensional physical system, there is a rich class of Lagrangians with
the property:
$\tfrac{\delta\,\mathbb{L}}{\delta\,q}=0$ $\Leftrightarrow$ $\{ \dot{x}=v\ \mbox{and}\ \dot{v}=-v\}$.

For more than one-dimensional ``\,world\,'' the situation becomes rather more complicated. The above simple partial
differential equation (\ref{integrator}) is replaced by a system of several partial differential equations and some
difficulties appear also on the ways $\Omega^\prime\rightsquigarrow\vartheta$ and $\vartheta\rightsquigarrow\mathbb{L}$.
But suppose there is a set of Lagrangians that generates the same classical dynamics. In this class one can introduce an
equivalence relation saying that $\mathbb{L}_1\sim\mathbb{L}_2$ $\Leftrightarrow$
$$
\bigl\{\theta_{\,\mathbb{L}_1}-\theta_{\,\mathrm{C}\mathbb{L}_2}= \,dF\,,\ \ \mbox{where}\ F\ \mbox{is a function on}\ TM\times\mathbb{R}\bigr\}
\ \Longleftrightarrow\
\bigl\{\mathbb{L}_1=\mathrm{C}\,\mathbb{L}_2+\tfrac{\partial\,F}{\partial\,t}+v^i\,\tfrac{\partial\,F}{\partial\,q^i}
\ \ \mbox{and}\ \ \tfrac{\partial\,F}{\partial\,v^i}=0\bigr\}\,.
$$
The equivalence relation $\sim$ has a very good motivation: whenever $\mathbb{L}_1\nsim\mathbb{L}_2$, then
$\Omega_1=d\theta_{\,\mathbb{L}_1}$ is equal to $\Omega_2=d\theta_{\,\mathbb{L}_2}$, but up to non-constant scaling
function $h$. This observation immediately implies that Lagrangians $\mathbb{L}_\alpha$, listed in the above one-dimensional
example, are not equivalent.

The non-equivalence of Lagrangians has strong physical consequences. The non-equivalent Lagrangians lead to
non-equivalent quantum mechanics (this is unrelated with the problem of the ordering), i.e. transition
amplitudes computed according to them are different, however, their classical limit is the same. This problem is called
a quantization ambiguity.

\section{Variational principle and ``\,umbilical\,'' string surfaces}

\begin{figure}[tbh]
\begin{center}
\psfrag{TQR}{\footnotesize{$TM\times\mathbb{R}$}}
\psfrag{thL}{\footnotesize{$\theta_{\,\mathbb{L}}$}}
\psfrag{t0}{$t_0$}
\psfrag{t1}{$t_1$}
\psfrag{TQt0}{\footnotesize{$TM\times\{t_0\}$}}
\psfrag{TQt1}{\footnotesize{$TM\times\{t_1\}$}}
\psfrag{R}{\footnotesize{$\mathbb{R}$}}
\psfrag{g}{$\gamma$}
\psfrag{q0}{\scriptsize{$(q_0,t_0)=\mathrm{fixed}$}}
\psfrag{q1}{\scriptsize{$(q_1,t_1)=\mathrm{fixed}$}}
\psfrag{wg}{$\Phi_{_{\delta W}}\gamma$}
\psfrag{dw}{\footnotesize{$\delta W\hspace{-0.5mm}\underset{\gamma}{\bigr|}$}}
\psfrag{ga}{$\tfrac{d\gamma}{d\tau}$}
\psfrag{s}{\footnotesize{$\tau$}}
\epsfxsize=12cm
\epsfbox{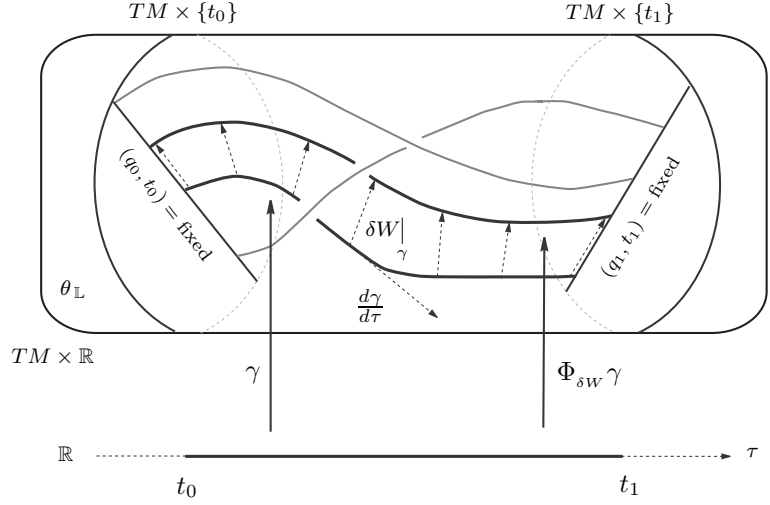}
\hspace{-3cm}
\caption{Schematic picture of the class $\mathscr{G}$. Any admissible curve should start at the $n$-dimensional
velocity-fiber $(q_0,t_0)=\mathrm{fixed}$ of $TM\times\{t_0\}$ and finish at the fiber $(q_1,t_1)=\mathrm{fixed}$ of
$TM\times\{t_1\}$.
The testing curve $\gamma$ and its variation $\Phi_{_{\delta W}}\gamma$ are both drawn by bold lines.}\label{obraz2}
\end{center}
\end{figure}

Let us shortly remind the reader with the standard variational approach. It is based on the Lepage one-form
$\theta_{\,\mathbb{L}}$ (in that case $\Omega=d\theta_{\,\mathbb{L}}$) and variation is carried over the curves in
the extended tangent bundle. Down-to-earth, let us choose two points (events) $(q_0,t_0)$ and
$(q_1,t_1)$ in the extended configuration space (space-time) $M\times\mathbb{R}$ and
consider the following class of admissible extended tangent bundle curves (see Figure \ref{obraz2}):
$$
\mathscr{G}:=\bigl\{
\gamma:\ \tau\in\langle t_0, t_1\rangle\ \mapsto \bigl(q=q(\tau), v=v(\tau), t=\tau\bigr)\in TM\times\mathbb{R}\,,\ \ \mbox{such that}\ \ q(\tau=t_0)=q_0\ \ \mbox{and}\ \ q(\tau=t_1)=q_1
\bigr\}\,.
$$
We shall see that the classical trajectory, specified before with the help of the null-spaces of $\Omega$,
belongs to the above class and extremizes the action
\begin{equation}
S:\ \gamma\ \mapsto\ S[\gamma]:=\int\limits_{\gamma}\theta_{\,\mathbb{L}}\,.
\end{equation}
Concisely, suppose that $\delta W=(\delta A)\partial_{t}+(\delta B^i)\partial_{q^i}+(\delta C^i)\partial_{v^i}$ is
an infinitesimal variational vector field in some small tubular neighborhood of the curve $\gamma$ (in order not to leave
with its flow $\Phi_{_{\delta W}}$ the class $\mathscr{G}$, we must require $\delta W$ to be tangential to
the edge submanifolds $(q_0,t_0)=\mathrm{fixed}$ and $(q_1,t_1)=\mathrm{fixed}$). Then, up to the first order in $\delta$,
we can write:
$$
\delta S[\gamma]=S[\Phi_{_{\delta W}}\gamma]-S[\gamma]=\int\limits_{\gamma}\mathcal{L}_{\delta W}(\theta_{\,\mathbb{L}})=\int\limits_{\gamma}d\theta_{\,\mathbb{L}}(\delta W,\,.\,)\,+\,
\underbrace{\theta_{\,\mathbb{L}}(\delta W)\hspace{-2mm}\underset{{{\gamma(t_1)}}}{\bigr|}\hspace{-2mm}-
\theta_{\,\mathbb{L}}(\delta W)\hspace{-2mm}\underset{{{\gamma(t_0)}}}{\bigr|}}_{\mathrm{each\ boundary\ term\ =\ 0}}=
\int\limits_{t_0}^{t_1}d\tau\,d\theta_{\,\mathbb{L}}\bigl(\delta W\hspace{-0.5mm}\underset{\gamma}{\bigr|}\hspace{-0.3mm},\tfrac{d\gamma}{d\tau}(\tau)\bigr)\,.
$$
The testing curve $\gamma$ is an extremal of $S$, if for any variation $\delta W$ it holds: $\delta S[\gamma]=0$.
This is only possible if for all $\tau\in\langle t_0,t_1\rangle$ the tangent vector
$$
\frac{d}{d\tau}\,\gamma\bigr|_{\gamma(\tau)}=\partial_{t}\bigr|_{\gamma(\tau)}
+\frac{dq^i}{d\tau}\,\partial_{q^i}\bigr|_{\gamma(\tau)}+\frac{dv^i}{d\tau}\,\partial_{v^i}\bigr|_{\gamma(\tau)}
$$
belongs to the kernel of $d\theta_{\,\mathbb{L}}$, quod erat demonstrandum.
Just for the completeness, the boundary terms drop out due to tangentiality of $\delta W$ to the edge submanifolds,
where the restriction of the Lepage one-form
$\theta_{\,\mathbb{L}}=\mathbb{L}\,dt+(\partial_{\,v^{i}}\,\mathbb{L})\,\bigl\{dq^i-v^i\,dt\bigr\}$ becomes trivial.

In the case when the distinguished two-form $\Omega$ does not possess a potential $\theta_{\,\mathbb{L}}$, we adopt
a new type of variational principle. As a byproduct, in the process of variation we get desired
classical trajectory plus something more.

\begin{figure}[tbh]
\begin{center}
\psfrag{q0t0}{\footnotesize{$(q_0,t_0)=\mathrm{fixed}$}}
\psfrag{q1t1}{\footnotesize{$(q_1,t_1)=\mathrm{fixed}$}}
\psfrag{t0s0}{\footnotesize{$(t_0,0)$}}
\psfrag{t0s1}{\footnotesize{$(t_0,1)$}}
\psfrag{t1s0}{\footnotesize{$(t_1,0)$}}
\psfrag{t}{\footnotesize{$\tau$}}
\psfrag{SI}{\large{$\boldsymbol{\Sigma}$}}
\psfrag{s}{\footnotesize{$\sigma$}}
\psfrag{t1s1}{\footnotesize{$(t_1,1)$}}
\psfrag{gr}{\footnotesize{$\gamma_{\mathrm{ref}}=\Sigma(\tau,\sigma=0)$}}
\psfrag{ge}{\footnotesize{$\gamma$}}
\psfrag{S1}{\footnotesize{$\Sigma(\tau,\sigma=1)$}}

\psfrag{lm}{\footnotesize{$\lambda_{0}$}}
\psfrag{lp}{\footnotesize{$\lambda_{1}$}}
\psfrag{TMR}{\footnotesize{$TM\times\mathbb{R}$}}
\psfrag{Om}{\footnotesize{$\Omega$}}

\psfrag{t=t0}{\tiny{$\tau=t_0$}}
\psfrag{t=t1}{\tiny{$\tau=t_1$}}
\psfrag{s=0}{\tiny{$\sigma=0$}}
\psfrag{s=1}{\tiny{$\sigma=1$}}

\epsfxsize=12cm
\epsfbox{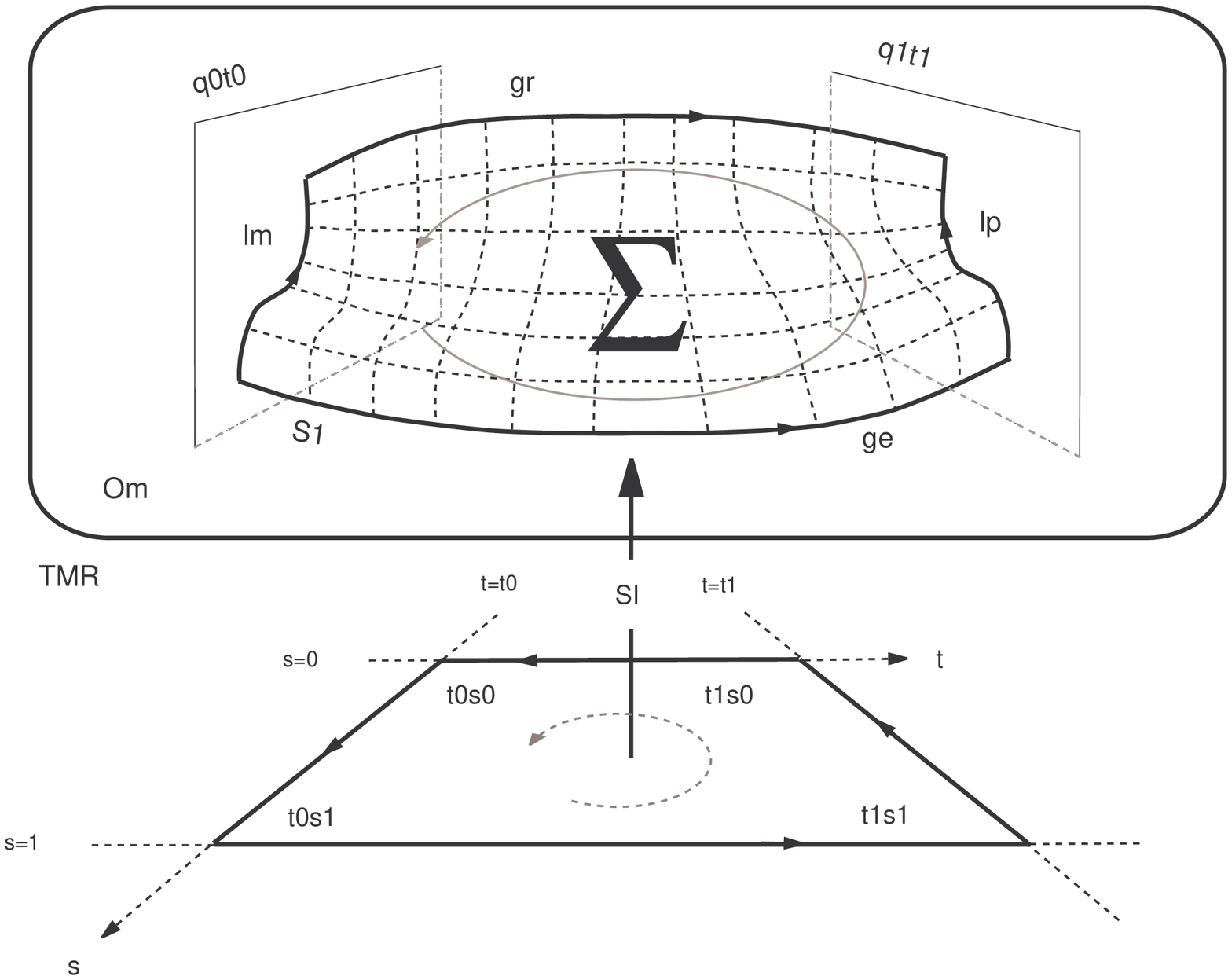}
\vspace{+1cm}
\caption{Oriented ``\,umbilical surface\,'' $\Sigma$ connects the reference curve $\gamma_{\mathrm{ref}}$ with the
``\,history\,''
$\gamma(\tau)=\Sigma(\tau,\sigma=1)$. Sideways boundary curves $\lambda_0(\sigma)=\Sigma(t_0,\sigma)$ and
$\lambda_1(\sigma)=\Sigma(t_1,\sigma)$ are located within the $n$-dimensional submanifolds
$(q_0,t_0)=\mathrm{fixed}$ and $(q_1,t_1)=\mathrm{fixed}$ of the extended tangent bundle.
On the figure, these edge submanifolds are schematically represented by two-\-dimensional ``\,D-branes.\,''}
\label{figure}
\end{center}
\end{figure}

Let us fix a reference curve $\gamma_{\mathrm{ref}}$ in the class $\mathscr{G}$ of all admissible curves and define
the space of its ``\,umbilical world-sheets\,'' (here $\tau$ is the time parameter as before, and $\sigma$ is a new
``\,worldsheet\,'' distance coordinate, see Figure \ref{figure}):
\begin{align*}
\mathscr{U}(\gamma_{\mathrm{ref}}):=
\Bigl\{
& \Sigma:\ (\tau,\sigma)\in\langle t_0, t_1\rangle\times\langle 0,1\rangle\ \mapsto \Sigma(\tau,\sigma)=\bigl(q=q(\tau,\sigma), v=v(\tau,\sigma), t=\tau\bigr)\in TM\times\mathbb{R}\,,
\ \mbox{such that}\\
& \mbox{for all values of the parameter}\ \sigma:\ q(\tau=t_0,\sigma)=q_0\,,\ q(\tau=t_1,\sigma)=q_1\,,\ \mbox{and}\ \Sigma(\tau,\sigma=0)=\gamma_{\mathrm{ref}}(\tau)
\Bigr\}\,.
\end{align*}

\noindent In the class of ``\,umbilical world-sheets\,'' a stationary surface
of the action
\begin{equation}\label{umbilical action}
S:\ \Sigma\ \mapsto\ S(\Sigma):=\int\limits_{\Sigma}\,\Omega
\end{equation}
satisfies:
\begin{equation}\label{variational eq.}
d\Omega\,\bigl(\tfrac{\partial}{\partial\tau}\,\Sigma(\tau,\sigma),\,\tfrac{\partial}{\partial\sigma}\,\Sigma(\tau,\sigma),\,.\,\bigr)=0\
 \ \ \ \ \ \ \ \ \ \mbox{and}\ \ \ \ \ \ \ \
 \Omega\,\bigl(\tfrac{\partial}{\partial\tau}\,\Sigma(\tau,\sigma=1),\,.\,\bigr)=0\,.
\end{equation}
To see this, let us consider in a small sandwich neighborhood of a testing surface $\Sigma\in\mathscr{U}(\gamma_{\mathrm{ref}})$
an infinitesimal variational vector field $\delta W$ (similarly as before, in order not to leave
with its flow
$\Phi_{_{\delta W}}$ the above ``\,umbilical\,'' class, we need, apart from tangentiality of $\delta W$ to the edge
submanifolds, also $\delta W|_{\gamma_{\mathrm{ref}}}=0$). Then, up to first order in $\delta$, we can write:

\begin{align*}
\delta S[\Sigma]& =S[\Phi_{_{\delta W}}\Sigma]-S[\Sigma]=\int\limits_{\Sigma}\mathcal{L}_{\delta W}(\Omega)=
\int\limits_{\Sigma}\delta W\lrcorner\,d\Omega+d\bigl(\delta W\lrcorner\,\Omega\bigr)=
\int\limits_{\Sigma}d\Omega(\delta W,.,.)+\int\limits_{\partial\,\Sigma}\Omega(\delta W,.)=\\
& =\int\limits_{t_0}^{t_1}\hspace{-1mm}d\tau\hspace{-1mm}\int\limits_{0}^{1}\hspace{-1mm}d\sigma \Bigl\{d\Omega\bigl(\delta W\hspace{-0.5mm}\underset{\Sigma}{\bigr|}\hspace{-0.5mm},\tfrac{\partial\,\Sigma}{\partial\sigma},\tfrac{\partial\,\Sigma}{\partial\tau}\bigr)\Bigr\}(\tau,\sigma)
+ \int\limits_{\gamma}\Omega(\delta W\hspace{-0.5mm}\underset{\gamma}{\bigr|}\hspace{-0.5mm},\,.)
- \underbrace{\int\limits_{\gamma_{\mathrm{ref}}}\Omega(\delta W\hspace{-2mm}\underset{\gamma_{\mathrm{ref}}}{\bigr|}\hspace{-2mm},\,.)
+\int\limits_{\lambda_1}\Omega(\delta W\hspace{-1mm}\underset{\lambda_1}{\bigr|}\hspace{-1mm},\,.)
-\int\limits_{\lambda_0}\Omega(\delta W\hspace{-1mm}\underset{\lambda_0}{\bigr|}\hspace{-1mm},\,.)}_{\mathrm{each\ boundary\ line\ integral\ is\ 0\ separately}}\,.
\end{align*}
The first underbraced integral is zero, because $\delta W|_{\gamma_{\mathrm{ref}}}=0$. The remaining two terms do not
contribute due to tangentiality of $\delta W$ to the edge ``\,D-brane\,'' submanifolds, where the restriction
$\Omega|_{\mathrm{D\mbox{-}branes}}$ equals to zero.
If the bulk term in $\delta S[\Sigma]$ obeys the first and the boundary term the second equation in
(\ref{variational eq.}) (we have used the abbreviation $\gamma(\tau)=\Sigma(\tau,\sigma=1)$), then $\delta S[\Sigma]$
is equal to zero for any variational vector field $\delta W$, quod erat demonstrandum.

The second equation in (\ref{variational eq.}) says that $\tfrac{\partial}{\partial\tau}\,\gamma(\tau)$
lies in the kernel of $\Omega$, therefore it determines the classical trajectory $\gamma_{\mathrm{class}}$ that connects
the space-time events $(q_0,t_0)$ and $(q_1,t_1)$.
Its genuineness is obvious, it does not depend on the chosen auxiliary reference curve
$\gamma_{\mathrm{ref}}\in\mathscr{G}$.

The complete solution of the first equation in (\ref{variational eq.}) is constrained by the boundary conditions
anchoring the stationary surface to the curves $\gamma_{\mathrm{ref}}$ and $\gamma_{\mathrm{class}}$.
Whether for any admissible reference curve $\gamma\in\mathscr{G}$ there exists a stationary ``\,umbilical\,'' surface
$\Sigma_{\mathrm{stat}}\in\mathscr{U}(\gamma)$ connecting it with the classical trajectory,
depends on the physical system under the consideration.
For example, for
conservative systems, where $d\Omega=d(d\theta_{\,\mathbb{L}})=0$, any world-sheet $\Sigma$
satisfying the boundary conditions: $\Sigma(\tau,\sigma=0)=\gamma(\tau)$ and $\Sigma(\tau,\sigma=1)=\gamma_{\mathrm{class}}(\tau)$
forms the extremal solution of (\ref{variational eq.}). Moreover, the prescription:
\begin{equation}\label{possible-action}
\gamma\ \mapsto\ \Sigma_{\mathrm{stat}}\ \mapsto\ S[\gamma]:=-\hspace{-2mm}\int\limits_{\Sigma_{\mathrm{stat}}}\hspace{-2mm}\Omega=-\hspace{-2mm}\int\limits_{\Sigma_{\mathrm{stat}}}\hspace{-2mm}d\theta_{\,\mathbb{L}}=-\hspace{-3mm}\int\limits_{\partial\,\Sigma_{\mathrm{stat}}}\hspace{-2mm}\theta_{\,\mathbb{L}}=
\int\limits_{\gamma}\theta_{\,\mathbb{L}}-\hspace{-2mm}\int\limits_{\gamma_{\mathrm{class}}}\hspace{-2mm}\theta_{\,\mathbb{L}}-
\underbrace{\int\limits_{\lambda_1}\theta_{\,\mathbb{L}}+\int\limits_{\lambda_0}\theta_{\,\mathbb{L}}}_{\mathrm{zero}}=
\int\limits_{\gamma}\theta_{\,\mathbb{L}}-\mathrm{C}
\end{equation}
is a meaningfull definition of the action for the history $\gamma$ (the constant $\mathrm{C}$ is the value of
the classical action on the trajectory $\gamma_{\mathrm{class}}$).

On the other hand, let us again focus on the one-dimensional particle driven by the force $\mathbb{Q}(x,v,t)=-v$. In this
case $d\Omega=-dv\,\wedge\,dx\,\wedge\,dt$ and a world-sheet $\Sigma$ given by the coordinate
functions $x=x(\tau,\sigma)$, $v=v(\tau,\sigma)$ and $t=\tau$ is stationary if
$$
0=d\Omega\,\bigl(\tfrac{\partial}{\partial\tau}\,\Sigma(\tau,\sigma),\,\tfrac{\partial}{\partial\sigma}\,\Sigma(\tau,\sigma),\,.\,\bigr)=
\frac{\partial x}{\partial\sigma}\,dv-\frac{\partial v}{\partial\sigma}\,dx+
\Bigl\{\frac{\partial x}{\partial\tau}\,\frac{\partial v}{\partial\sigma}-\frac{\partial v}{\partial\tau}\,\frac{\partial x}{\partial\sigma}\Bigr\}\,dt
\ \Longleftrightarrow\
\begin{cases} x(\tau,\sigma)=x(\tau,\sigma=0) \\ v(\tau,\sigma)=v(\tau,\sigma=0) \end{cases}
$$
Whenever $\Sigma(\tau,\sigma=0)$ is not equal to $\Sigma(\tau,\sigma=1)$, there does not exist a non-degenerate
solution of (\ref{variational eq.}), i.e. extremal ``\,umbilical\,'' string satisfying the given boundary
conditions $\Sigma(\tau,\sigma=0)=\gamma$ and $\Sigma(\tau,\sigma=1)=\gamma_{\mathrm{class}}$ is missing.
If we accept in the ``\,umbilical\,'' class $\mathscr{U}(\gamma)$ also a degenerate surface (this one
is shrunk just to the reference curve), then for the special gauge $\gamma=\gamma_{\mathrm{class}}$ we
get the solution of the full system (\ref{variational eq.}) in the relatively simple form
$\Sigma_{\mathrm{stat}}(\tau,\sigma)=\gamma_{\mathrm{class}}(\tau)$.

All of this explains in advance why it is impossible to apply the assignment (\ref{possible-action}):
$\gamma\mapsto\Sigma_{\mathrm{stat}}\mapsto S[\gamma]$ as the universal principle. This is the main difference compared to
the result of Mari\'{a}n Fecko's paper\cite{fecko2}, in which he analyzed a similar problem in the context of a
variational principle for the Nambu mechanics.

We will see that although the value of the classical action $S[\gamma]$ is missing, it is still
possible to perform a quantization of the theory in terms of the two-form $\Omega$.

\section{Quantization: Path \emph{versus} Surface Integral }

In the previous sections we have observed that the classical evolution is completely described when
integrating null spaces $\ell$ of the two-form $\Omega$. So we could claim:
``\,classical mechanics is only $\Omega$-sensitive.'' Everything else is a bonus only valid in special cases.
We were being impractical not to use the (local) potential $\theta_{\,\mathbb{L}}$ or the Lagrangian $\mathbb{L}$, which
would enable us to investigate the invariants and/or conserved quantities. But the physical principles are constituted
over the equations of motion, not over the Lagrangian or Hamiltonian themselves.

On the other hand, it seems that the standard quantum mechanics is rather $\theta_{\,\mathbb{L}}$ (or, if you
wish, $\omega_{\,\mathbb{H}}$)-sensitive. The most impressive way how to relate classical and quantal lies in
the Feynman path-integral approach.

According to the Feynman prescription\cite{feynman-hibbs}${}^,$\cite{faddeev-slavnov} the probability amplitude of the
transition of the system from the space-time configuration $(q_0,t_0)$ to $(q_1,t_1)$ is expressed as
follows:
\begin{equation}\label{FeynmanI}
\mathbf{A}(q_0,t_0;q_1,t_1)\ \propto\,
{\Large{\int\limits_{\mathscr{G}}}} [\mathscr{D}\gamma]\,\exp{\Bigl\{\frac{i}{\hbar}\int\limits_{\gamma}\,\theta_{\,\mathbb{L}}\Bigr\}}\,.
\end{equation}
The ``\,path-summation\,'' here is taken over the class $\mathscr{G}$ of all admissible curves in $TM\times\mathbb{R}$
as it is drawn in Figure \ref{obraz2}.
The exponent in (\ref{FeynmanI}) is the standard integral of the one-form $\theta_{\,\mathbb{L}}$ carried over
the extended tangent bundle curve $\gamma$. The question about the ``\,measure\,'' $[\mathscr{D}\gamma]$ and
the proper normalization of the probability amplitude $\mathbf{A}$ are subject to our discussion in the next
section. Let us remind the reader that the probability amplitude formula
(\ref{FeynmanI}) is used less frequently than its phase space version. When expressing generalized velocities
in $\theta_{\,\mathbb{L}}$ in terms of generalized momenta we get $\mathbf{A}$ as a functional integral in
the extended phase space $T^*M\times\mathbb{R}$:
$$
\mathbf{A}(q_0,t_0;q_1,t_1)\ =\,
\int\limits_{\widetilde{\mathscr{G}}} [\mathscr{D}\widetilde{\gamma}]\,\exp{\Bigl\{\frac{i}{\hbar}\int\limits_{\widetilde{\gamma}}\,\omega_{\,\mathbb{H}}\Bigr\}}\,,
\ \ \mbox{where one can formally set}\ \
[\mathscr{D}\widetilde{\gamma}]=\dfrac{dp_{{t_0}}}{2\pi}\,\prod\limits_{\overset{t\in(t_0,t_1)}{{}}}\dfrac{dp_{_t}\,dq_{_t}}{2\pi}\,.
$$
The bunches of curves $\gamma$ and
$\widetilde{\gamma}$ that enter the functional integrations are connected by the same Legendre transformation
$(q,v,t,\mathbb{L})\mapsto(q,p=\tfrac{\partial\mathbb{L}}{\partial v},t,\mathbb{H}=pv-\mathbb{L})$ as the one-forms
$\theta_{\,\mathbb{L}}$ and $\omega_{\,\mathbb{H}}$.

The above mentioned sensitiveness of quantum mechanics on the one-form $\theta_{\,\mathbb{L}}$ and/or $\omega_{\,\mathbb{H}}$
is evident.
In what follows, we propose some modifications leading to the replacement of $\theta_{\,\mathbb{L}}$
by the two-form $\Omega$. This will enable us to ``\,quantize\,'' also dissipative systems. In the special conservative
case, our prescription will be equivalent to the Feynman's.

Our main trick is a simple rearrangement based on the Stokes theorem. Down-to-earth,
in the class $\mathscr{G}$ that enters the ``\,path-summation\,'' in (\ref{FeynmanI}), there is one specially
distinguished curve, the classical trajectory\footnote{We optimistically propose that
solutions of the equations of motion might be ``\,inverted\,'' on relatively broad time interval, i.e. that from given
position at the final time we would be able to adjust the initial velocity in such a way that the system will evolve
uniquely into the prescribed endpoint.} $\gamma_{\mathrm{class}}$.
Using it, we get for any other $\gamma$ within this class an oriented 1-cycle:
$$
\partial\,\Sigma:=\gamma+\lambda_1-\gamma_{\mathrm{class}}-\lambda_0\,.
$$
Here $\lambda_0$ and $\lambda_1$ are arbitrarily chosen curves within the submanifolds of $TM\times\mathbb{R}$
$(q_0,t_0)=\mathrm{fixed}$ and $(q_1,t_1)=\mathrm{fixed}$ that join the initial and final points of $\gamma$ and
$\gamma_{\mathrm{class}}$, respectively (the choice $\gamma_{\mathrm{ref}}=\gamma_{\mathrm{class}}$ in Figure
\ref{figure} provides a correct picture of this situation). Since the restriction
of $\theta_{\,\mathbb{L}}$ (and also of the distinguished two-form $\Omega$) to any of these edge ``\,D-brane\,''
submanifolds is trivial, we can write:
\begin{equation}\label{I}
\int\limits_{\gamma}\,\theta_{\,\mathbb{L}}\,-\hspace{-0.8mm}\int\limits_{\gamma_{\mathrm{class}}}\hspace{-0.8mm}\theta_{\,\mathbb{L}}\,
+\underbrace{\int\limits_{\lambda_1}\theta_{\,\mathbb{L}}-\int\limits_{\lambda_0}\theta_{\,\mathbb{L}}}_{\mathrm{zero}}=\,
\int\limits_{\partial\,\Sigma}\,\theta_{\,\mathbb{L}}\,=\,\int\limits_{\Sigma}\,d\theta_{\,\mathbb{L}}\,,
\ \ \mbox{where}\ \ \Sigma\in\mathscr{U}(\gamma_{\mathrm{class}})\ \ \mbox{and}\ \ \partial\,\Sigma=\gamma+\lambda_1-\gamma_{\mathrm{class}}-\lambda_0\,.
\end{equation}
Let us remind the reader that
\begin{itemize}
\item[$\circ$] the second term on the left hand side of (\ref{I}) is just the value of the classical action on the curve $\gamma_{\mathrm{class}}$
\item[$\circ$] the existence of the ``\,umbilical\,'' string $\Sigma$ that connects the given curve $\gamma$ with $\gamma_{\mathrm{class}}$
         is determined by topological properties\footnote{Topological properties we are talking about are ``\,measured\,''
         by the fundamental group $\Pi_1(TM\times\mathbb{R})$. For obvious reasons we are cowardly skipping off any
         discussion of the quantization in topologically nontrivial cases.} of
         $TM\times\mathbb{R}$, e.g. when $TM\times\mathbb{R}$ is simply-connected, then any
         1-cycle $\partial\,\Sigma$ is at the same time a 1-boundary of some 2-chain $\Sigma$
\end{itemize}
Motivated by the trick (\ref{I}), encouraged by the sentence from Richard Feynman's
thesis:\cite{feynman} \emph{...\,the central mathematical concept is the analogue of the action in classical mechanics.
It is therefore applicable to mechanical systems whose equations of motion cannot be put into Hamiltonian form. It is
only required that some sort of least action principle be available\,...} and inspired by the variational principle
(\ref{umbilical action}), we propose a generalization of the Feynman's probability amplitude formula in the following way:
\begin{equation}\label{FeynmanII}
\mathbf{A}(q_0,t_0;q_1,t_1)\ \propto\,
\exp{\Bigl\{\frac{i}{\hbar}\,S_{\mathrm{class}}\Bigr\}}\ \int\limits_{\mathscr{U}} [\mathscr{D}\Sigma]\,\exp{\Bigl\{\frac{i}{\hbar}\int\limits_{\Sigma}\,\Omega\Bigr\}}\,.
\end{equation}
Here the ``\,surface-summation\,'' is taken over the class $\mathscr{U}\equiv\mathscr{U}(\gamma_{\mathrm{class}})$ of all
admissible ``\,umbilical\,'' world-sheets with the reference curve $\gamma_{\mathrm{class}}$.
Using the formula (\ref{Omega}) for the distinguished two-form $\Omega$ we can write:
$$
\int\limits_{\Sigma}\,\Omega\,=\int\limits_{\partial\,\Sigma}\,\bigl\{\mathbb{T}\,dt+(\partial_{v^{i}}\,\mathbb{T})\,\{dq^i-v^i\,dt\}\bigr\}
+\int\limits_{\Sigma}\,\{\mathbb{Q}\,_i\,dq^i\}\wedge\,dt\,\equiv
\int\limits_{\partial\,\Sigma}\,\theta_{\,\mathbb{T}}+\int\limits_{\Sigma}\,\{\mathbb{Q}\,_i\,dq^i\}\wedge\,dt
\,.
$$
The first integral term is obviously independent of the choice of the sideways boundary curves $\lambda_0$ and
$\lambda_1$ in $\partial\,\Sigma=\gamma+\lambda_{1}-\gamma_{\mathrm{class}}-\lambda_{0}$. Moreover, we can split
the ``\,surface-summation\,'' carried out in (\ref{FeynmanII}) in the following way:
$$
\int\limits_{{\mathscr{U}}}[\mathscr{D}\Sigma]\,=\int\limits_{\mathscr{G}}[\mathscr{D}\gamma]\ \biggl\{\,\int\limits_{\{\Sigma_\gamma\}}[\mathscr{D}\Sigma_{\gamma}]\,\biggr\}\,,
$$
i.e. first we pick out the boundary curve $\gamma\in\mathscr{G}$, and then we perform the ``\,summation\,'' over the
subset
$$
\{\Sigma_\gamma\}:=
\bigl\{
\Sigma_{\gamma}\in\mathscr{U}\,,\ \mbox{such that}\ \Sigma_{\gamma}(\tau,\sigma=1)=\gamma(\tau)
\bigr\}\subset\mathscr{U}\,,
$$
which contains all ``\,umbilical\,'' surfaces that are anchored to the fixed curves $\gamma_{\mathrm{class}}$ and $\gamma$.
After doing this, we get (\ref{FeynmanII}) in the equivalent form:
$$
\mathbf{A}(q_0,t_0;q_1,t_1)\propto\,
\exp{\Bigl\{\frac{i}{\hbar}\,S_{\mathrm{class}}\Bigr\}}\int\limits_{\mathscr{G}}
[\mathscr{D}\gamma]\exp{\biggl\{\frac{i}{\hbar}\Bigl\{\int\limits_{\gamma}-\int\limits_{\gamma_{\mathrm{class}}}\Bigr\}\theta_{\,\mathbb{T}}\biggr\}}
\times\Biggl\{\hspace{1mm}\int\limits_{\{\Sigma_{\gamma}\}}[\mathscr{D}\Sigma_\gamma]\exp{\biggl\{\frac{i}{\hbar}\int\limits_{\Sigma_\gamma}\{\mathbb{Q}\,_i\,dq^i\}\wedge\,dt\biggr\}}\Biggr\}\,.
$$
In the case of conservative forces $\{\mathbb{Q}\,_i\,dq^i\}\wedge\,dt=-d\theta_{\,\mathbb{U}}=-d\bigl\{\mathbb{U}\,dt+(\partial_{v^{i}}\,\mathbb{U})\,\{dq^i-v^i\,dt\}\bigr\}$,
the surface integral in the last exponent of the above formula is again only boundary sensitive quantity. Therefore
$$
\mathbf{A}(q_0,t_0;q_1,t_1)\ \propto\,
\exp{\Bigl\{\frac{i}{\hbar}\,S_{\mathrm{class}}\Bigr\}}\ \int\limits_{\mathscr{G}}
[\mathscr{D}\gamma]\,\exp{\biggl\{\frac{i}{\hbar}\Bigl\{\int\limits_{\gamma}-\int\limits_{\gamma_{\mathrm{class}}}\Bigr\}\,\bigl(\theta_{\,\mathbb{T}}-\theta_{\,\mathbb{U}}\bigr)\biggr\}}\,\times\,\mathrm{Vol}_{\gamma}\,,
$$
where we have adopted the abbreviated notation
$$
\mathrm{Vol}_{\gamma}\,=\int\limits_{\{\Sigma_\gamma\}}\,[\mathscr{D}\Sigma_\gamma]=\,\mbox{the
``\,number\,'' of the surfaces containing}\ \gamma\ \mbox{and}\ \gamma_{\mathrm{class}}\ \mbox{as the subboundaries}\,.
$$
Suppose there are no topological obstructions on the side of $TM\times\mathbb{R}$, i.e. that all admissible $\gamma$'s are
homoto\-pi\-cally equivalent. Then the factor $\mathrm{Vol}_\gamma$ is $\gamma$-independent, and it can be dropped
as an infinite constant by normalization. Moreover, the factor $S_{\mathrm{class}}$ cancels the
integral $\int(\theta_{\,\mathbb{T}}-\theta_{\,\mathbb{U}})$ over the reference curve $\gamma_{\mathrm{class}}$.
Thus, in the case of conservative forces the formula (\ref{FeynmanII}) reduces precisely to (\ref{FeynmanI}),
quod erat demonstrandum.

There is still one open point, namely we have to say what exactly we mean by the classical action $S_{\mathrm{class}}$
in (\ref{FeynmanII}). The distinguished two-form $\Omega$ specifies classical trajectory $\gamma_{\mathrm{class}}$.
It is defined directly in terms of acting forces. Some of them could be potential-generated and therefore $\Omega$ can be
split (at least locally) into the two parts:
$$
\Omega=\bigl\{\mathbb{Q}^{^{\mathrm{diss}}}_{\,i}\,dq^i\bigr\}\wedge\,dt+d\Bigl\{\bigl(\mathbb{T}-\mathbb{U}\bigr)\,dt+\partial_{\,v^{i}}\bigl(\mathbb{T}-\mathbb{U}\bigr)\,\bigl\{dq^i-v^i\,dt\bigr\}\Bigr\}
=:\,\Omega_{\mathrm{diss}}+\Omega_{\mathrm{cons}}\,.
$$
The first part, called dissipative, contains all the non-potential-generated forces and therefore
$d\Omega_{\mathrm{diss}}\neq 0$. The remaining closed term
$\Omega_{\mathrm{cons}}=d\{\theta_{\,\mathbb{T}}-\theta_{\,\mathbb{U}}\}$ is a conservative part of $\Omega$.
The classical action entering the amplitude formula (\ref{FeynmanII}) is defined as follows:
\begin{equation}\label{classical action}
S_{\mathrm{class}}:=\hspace{-1mm}\int\limits_{\gamma_{\mathrm{class}}}\hspace{-1mm}\bigl(\theta_{\,\mathbb{T}}-\theta_{\,\mathbb{U}}\bigr)\,.
\end{equation}

We described above all objects that are necessary for the computation of the probability amplitude. It remains to
give some nontrivial example demonstrating the functionality of (\ref{FeynmanII}) and then
open the discussion.
Let us start from backward, with the discussion, and postpone the example to the next paragraph:
\begin{itemize}
\item[$\circ$] to see the correctness of the classical limit and the exceptionality of the classical history in between
               $(q_0,t_0)$ and $(q_1,t_1)$, one needs to recollect the above ``\,umbilical\,'' variational principle
               (\ref{umbilical action}); the stationary solution $\Sigma_{\mathrm{stat}}$ of variational equations
               (\ref{variational eq.}) in the considered ``\,umbilical\,'' class $\mathscr{U}$ corresponds to
               $\gamma_{\mathrm{class}}$,
\item[$\circ$] here, to be able to talk about the quantum probability amplitudes, one needs to know the solution of
               the classical equations of motion with the given initial condition; in the standard approach, the classical
               solution is not necessary for the quantization, it rather
               appears as the saddle point dominating the amplitude in the limit $\hbar\rightarrow 0$,
\item[$\circ$] the mathematically delicate question of the surface functional measure will not be discussed here; our
               understanding of this problem will be demonstrated on a special example in the next section,
\item[$\circ$] composition of the transition amplitudes we are accustomed to from the standard quantum mechanics:
               $$
               \mathbf{A}(q_0,t_0;q_1,t_1)=\int\limits_{M}\,dq\,\mathbf{A}(q_0,t_0;q,t)\,\mathbf{A}(q,t;q_1,t_1)
               $$
               does not work if $d\Omega\neq 0$, i.e. the dissipative time evolution does not satisfy a groupoid
               composition rule: $U(t_0,t_1)= U(t_0,t)\circ U(t,t_1)$;
\end{itemize}

\section{An example: quantum mechanics with friction}

Let us focus on the quantization of dynamics of a unit mass particle moving in $M=\mathbb{R}[x]$, which is driven by
the conservative force $\mathbb{F}=-\tfrac{d}{dx}\mathbb{U}(x)$ along with the friction $\mathbb{Q}^{^\mathrm{diss}}=-\kappa\, v$.
The extended tangent bundle for this situation corresponds to an ordinary three-dimensional Cartesian space
$\mathbb{R}^3[x,v,t]$, and the distinguished two-form takes the simple form
$\Omega=-\kappa\, v\,dx\,\wedge\, dt+d\,\bigl\{v\,dx-\{\tfrac{1}{2}\,v^2+\mathbb{U}(x)\}\,dt\bigr\}$.

Our aim is to evaluate the transition amplitude as a function of the initial and final events. To keep
the better track of the forthgoing calculation and also of its frictionless limit, we maintain the explicit
dependence on the parameter $\kappa$.
Suppose we have chosen a solution $\gamma_{\mathrm{class}}(\tau)=\bigl(x_{\mathrm{class}}(\tau),v_{\mathrm{class}}(\tau)=\dot{x}_{\mathrm{class}}(\tau),t=\tau\bigr)$
of the Newton-Lagrange equation of motion:
$$
\ddot{x}=\mathbb{F}(x)-\kappa\,\dot{x}\,,\ \ \ \ \ \mbox{which satisfies:}\ \ x_{\mathrm{class}}(t_0)=q_0\ \ \mbox{and}\ \ x_{\mathrm{class}}(t_1)=q_1\,.
$$
Direct application of the formula (\ref{FeynmanII}) leads then to the following expression:
\begin{equation}\label{QM}
\mathbf{A}(q_0,t_0;q_1,t_1)\propto\
\exp{\Bigl\{\frac{i}{\hbar}\,S_{\mathrm{class}}\Bigr\}}
\int\limits_{\mathscr{U}}[\mathscr{D}\Sigma]
\exp{\biggl\{\frac{i}{\hbar}\Bigl\{\int\limits_{\partial\,\Sigma}\,\bigl\{v\,dx-\bigl\{\tfrac{1}{2}\,v^2+\mathbb{U}(x)\bigr\}\,dt\bigr\}-\kappa\int\limits_{\Sigma} v\,dx\wedge dt \Bigr\}\biggr\}}\,.
\end{equation}
The meaning of the first exponent is clear from (\ref{classical action}). To evaluate the remaining world-sheet
functional integral, let us introduce an auxiliary nodal set:
$$
\{(a,b)\}_{\mathrm{nodes}}:=\bigl\{(t_0+a\,\varDelta,0+b\,\varepsilon)\in\mathbb{R}^2[\tau,\sigma]\,;\ \mbox{time index $a$ runs from $0$ to $K$ and distance index $b$ from $0$ to $L$}\bigr\}\,,
$$
which splits the underlying parametric space $\langle t_0,t_1\rangle\times\langle 0,1\rangle\subset\mathbb{R}^2[\tau,\sigma]$
into infinitesimal rectangular tiles, each of which encloses the area
$\varDelta\cdot\varepsilon=\tfrac{t_1-t_0}{K}\cdot\tfrac{1-0}{L}$ (at the end, the numbers $K$ and $L$ will be send to
infinity).
\begin{figure}[tbh]
\begin{center}
\psfrag{D}{\footnotesize{$\varDelta$}}
\psfrag{e}{\footnotesize{$\varepsilon$}}
\psfrag{(a,b)}{\footnotesize{$(a,b)$}}
\psfrag{t+aD}{\tiny{$t_0+a\,\varDelta$}}
\psfrag{s+be}{\tiny{$0+b\,\varepsilon$}}
\psfrag{a=0}{\tiny{$a=0$}}
\psfrag{a=1}{\tiny{$a=1$}}
\psfrag{a=K}{\tiny{$a=K$}}
\psfrag{b=0}{\tiny{$b=0$}}
\psfrag{b=1}{\tiny{$b=1$}}
\psfrag{b=L}{\tiny{$b=L$}}

\psfrag{t=0}{\tiny{$\tau=t_0$}}
\psfrag{t=1}{\tiny{$\tau=t_1$}}
\psfrag{s=0}{\tiny{$\sigma=0$}}
\psfrag{s=1}{\tiny{$\sigma=1$}}

\epsfxsize=12cm
\epsfbox{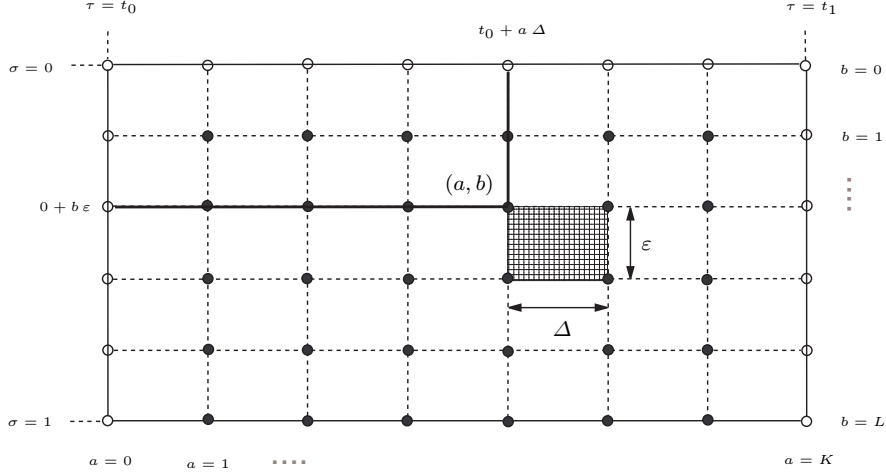}
\vspace{+1cm}
\caption{Schematic picture of the nodal grid, points marked with empty circles are constrained by (\ref{cases}).}
\label{grid}
\end{center}
\end{figure}

\noindent After that any ``\,umbilical\,'' string
$$
\Sigma:\ \mathbb{R}^2[\tau,\sigma]\rightarrow\mathbb{R}^3[x,v,t]\,,\ \ \ \ (\tau,\sigma)\mapsto\Sigma(\tau,\sigma)=\bigl(x(\tau,\sigma),v(\tau,\sigma),t(\tau,\sigma)=\tau\bigr)
$$
can be discretized by evaluating its coordinate functions at the nodes of the considered grid $\{(a,b)\}_{\mathrm{nodes}}$,
i.e.
$$
\Sigma: \{(a,b)\}_{\mathrm{nodes}} \rightarrow \bigl\{\Sigma(a,b)=\bigl(x_{(a,b)},v_{(a,b)},t_0+a\,\varDelta\bigr)\bigr\}\,.
$$
Here, to keep the ensemble $\{\Sigma(a,b)\}\subset\mathbb{R}^3[x,v,t]$ within the considered ``\,umbilical\,''
class $\mathscr{U}$, we must impose the following simple constraints:
\begin{equation}\label{cases}
\begin{cases}
\forall\ a=0,\dots,K & x_{(a,0)}=x_{\mathrm{class}}(t_0+a\,\varDelta)\,,\ v_{(a,0)}=v_{\mathrm{class}}(t_0+a\,\varDelta)\ \Leftrightarrow\ \Sigma(\tau,\sigma=0)=\gamma_{\mathrm{class}}(\tau)\,,\\
\forall\ b=1,\dots,L & x_{(0,b)}=q_0\,,\ x_{(K,b)}=q_1\ \Leftrightarrow\ \Sigma(\tau=t_0,\sigma)\ \mbox{and}\ \Sigma(\tau=t_1,\sigma)\ \subset\ \mbox{to the ``\,D-branes\,''}\,.
\end{cases}
\end{equation}
Therefore, formally, the functional integral over all possible string configurations is a formal limit of the ordinary
multiple integral, which is taken over all unconstraint variables in the discretized ensemble $\{\Sigma(a,b)\}$, i.e.:
$$
\int\limits_{\mathscr{U}}[\mathscr{D}\Sigma]:=\lim\limits_{{K\rightarrow\infty \atop \,L\rightarrow\infty}}\
\biggl\{\int\limits_{-\infty}^{+\infty}\cdots\int\limits_{-\infty}^{+\infty}\ \,
\prod\limits_{a=1}^{K-1}\prod\limits_{b=1}^{L}\,dx_{(a,b)}\,dv_{(a,b)}\,dv_{(0,b)}\,dv_{(K,b)}\biggr\}\,.
$$
When step-wisely discretizing the integrals in the exponent of (\ref{QM}), taking into account the constraints
(\ref{cases}), we get for the bulk term (all is done with respect to the chosen orientation of the
``\,umbilical\,'' world-sheet $\Sigma$):
\begin{align*}
& \int\limits_{\Sigma}\,v\,dx\wedge dt=\int\limits_{t_0}^{t_1} d\tau \int\limits_{0}^{1}\,d\sigma\,\bigl\{v(\tau,\sigma)\,\tfrac{\partial x}{\partial \sigma}(\tau,\sigma)\bigr\}=
\lim\limits_{{K\rightarrow\infty \atop \,L\rightarrow\infty}}\ \sum\limits_{a=0}^{K-1}\sum\limits_{b=0}^{L-1}\,\varDelta\cdot\varepsilon\ \Bigl\{v_{(a,b)}\ \frac{x_{(a,b+1)}-x_{(a,b)}}{\varepsilon}\Bigr\}=\\
& = \lim\limits_{{K\rightarrow\infty \atop \,L\rightarrow\infty}}\biggl\{\,
\sum\limits_{a=1}^{K-1}\sum\limits_{b=1}^{L-1}\,\varDelta\,v_{(a,b)}\,\{x_{(a,b+1)}-x_{(a,b)}\}+
\sum\limits_{a=0}^{K-1}\,\varDelta\,v_{(a,0)}\,x_{(a,1)}\biggr\}
-\int\limits_{t_0}^{t_1}\,d\tau\,\bigl\{ v_{\mathrm{class}}(\tau)\,x_{\mathrm{class}}(\tau)\bigr\}
\end{align*}
and similarly for the boundary term:
\begin{align*}
& \int\limits_{\partial\,\Sigma}\,\bigl\{v\,dx-\bigl\{\tfrac{1}{2}\,v^2+\mathbb{U}(x)\bigr\}\,dt\}=
\hspace{-3mm}\int\limits_{\Sigma(\tau,\sigma=1)}\hspace{-2mm}\bigl\{v\,dx-\bigl\{\tfrac{1}{2}\,v^2+\mathbb{U}(x)\bigr\}\,dt\}\ -
\hspace{-3mm}\int\limits_{\Sigma(\tau,\sigma=0)}\hspace{-2mm}\bigl\{v\,dx-\bigl\{\tfrac{1}{2}\,v^2+\mathbb{U}(x)\bigr\}\,dt\}=\\
& =\lim\limits_{{K\rightarrow\infty \atop {}}}\sum\limits_{a=0}^{K-1}\Bigl\{v_{(a,L)}\{x_{(a+1,L)}-x_{(a,L)}\}-\varDelta\,\bigl\{\tfrac{1}{2}\bigl(v_{(a,L)}\bigr)^2+\mathbb{U}\bigl(x_{(a,L)}\bigr)\bigr\}\Bigr\}
-\int\limits_{t_0}^{t_1}\,d\tau\,\bigl\{\tfrac{1}{2}\,\bigl(v_{\mathrm{class}}(\tau)\bigr)^2-\mathbb{U}\bigl(x_{\mathrm{class}}(\tau)\bigr)\bigr\}\,.
\end{align*}
Putting everything together, integrating over all variables apart $\{x_{(a,L)}\}$ and returning back to the continuum
limit, we get the following expression for the transition amplitude:\footnote{To arrive to the above formula, one should handily employ, when performing the routine integrations over the velocities,
the spectral form of $\delta$-function and the Gauss-Fermi integral. After that, the remaining integrals carried out over
the ensemble $\{x_{(a,b)},\,a=1,\dots,K-1,\,b=1,\dots,L-1\}$ become trivial.}
\begin{equation}\label{QMII}
\mathbf{A}(q_0,t_0;q_1,t_1)\propto\
\exp{\Bigl\{\frac{i}{\hbar}\int\limits_{t_0}^{t_1}d\tau\bigl\{\kappa\,v_{\mathrm{class}}\,x_{\mathrm{class}}\bigr\}\Bigr\}}
\int\,[\mathscr{D}x(\tau)]
\exp\biggl\{\frac{i}{\hbar}\int\limits_{t_0}^{t_1}d\tau\bigl\{\tfrac{1}{2}\,\dot{x}^2-\mathbb{U}(x)-\kappa\,x\,v_{\mathrm{class}}\bigr\}\biggr\}\,.
\end{equation}
The phase factor in front of (\ref{QMII}) is coming from the definition of the classical action $S_{\mathrm{class}}$ and
from the world-sheet functional integration. The second term is the standard Feynman path integral, which is taken over
the histories $\{\tau\mapsto \bigl(x(\tau):=x(\tau,\sigma=1),t=\tau\bigr)\}$ in the extended configuration space
$M\times\mathbb{R}[t]$.
However, in comparison with the standard formula, a new term appears here. It is an external source generated by the classical
velocity $v_{\mathrm{class}}$. Its presence has an important consequence, namely, it guarantees that the classical
solution $x_{\mathrm{class}}(\tau)$ is the stationary curve of the considered functional.
Moreover, further inspection of (\ref{QMII}) shows
that if the friction parameter $\kappa$ tends to zero, the standard propagator in the potential
$\mathbb{U}(x)$ is recovered.\\

Let us remind the reader how to treat ugly infinite constants emerging in the functional integration. If the
entering infinities are functionally independent of the coordinates of space-time events, then one can easily neglect
them. The only important term inside the functional integral is the phase factor, which depends on coordinates of
$(q_0,t_0)$ and $(q_1,t_1)$, i.e. we need to seize the following quantity:
$$
\mathbf{A}(q_0,t_0;q_1,t_1)\propto\ \exp{\bigl\{\frac{i}{\hbar}\,\Phi(q_0,t_0;q_1,t_1)\bigr\}}\,,
$$
anything else is just an inherited rudiment. The proper normalization of the amplitude
$\mathbf{A}(q_0,t_0;q_1,t_1)$ is dictated by its physical meaning. The square of its absolute value answers the question
about the probability density to observe a particle in a sufficiently small neighborhood of the configuration
$(q_1,t_1)$, when before it was observed in a neighborhood of the space-time position $(q_0,t_0)$. This implies desired
normalization conditions (since we are dealing with the space-time continuum, the normalization to $\delta$-function
should be employed):
\begin{align*}
\bigl\{\mbox{if}\ \ t_1\rightarrow t_0\ \Longrightarrow\ \mathbf{A}(q_0,t_0;q_1,t_1)\rightarrow\delta(q_1-q_0)\bigr\}
&\ \Longleftrightarrow\ \bigl\{\mbox{{at the time $t_0$ system occupies definite position $q_0$}}\bigr\}\\
\int\limits_{-\infty}^{+\infty}dq_1\,\mathbf{A}(q_0,t_0;q_1,t_1)\,\boldsymbol{\mathbf{A}^*}(q_0^\prime,t_0;q_1,t_1)=\delta(q_0^\prime-q_0)
&\ \Longleftrightarrow\ \bigl\{\mbox{{total probability is conserved, evolution is unitary}}\bigr\}
\end{align*}\\

\noindent Having everything at hand, let us compute the normalized probability amplitude with the presence
of friction in the cases when $\mathbb{U}(x)=0$ (free particle) and $\mathbb{U}(x)=\tfrac{1}{2}\,\omega\,x^2$
(linear harmonic oscillator).

\subsection{Free particle with damping}

The classical trajectory $x_{\mathrm{class}}(\tau)$ with prescribed endpoints that satisfies the dynamical equation
$\ddot{x}=-\kappa\,\dot{x}$,
is given as follows:
$$
x_{\mathrm{class}}(\tau)=\frac{q_1\,\mathrm{e}^{\kappa t_1}-q_0\,\mathrm{e}^{\kappa t_0}}{\mathrm{e}^{\kappa t_1}-\mathrm{e}^{\kappa t_0}}
+\frac{q_1-q_0}{\mathrm{e}^{-\kappa t_1}-\mathrm{e}^{-\kappa t_0}}\ \mathrm{e}^{-\kappa\tau}\,,
\ \ \ v_{\mathrm{class}}(\tau)=\frac{d}{d\tau}x_{\mathrm{class}}(\tau)\ \ \ \mbox{and}\ \ \
\begin{cases}
x_{\mathrm{class}}(t_0)=q_0\,, \\
x_{\mathrm{class}}(t_1)=q_1\,.
\end{cases}
$$
The only difficulty that comes from (\ref{QMII}), is the path integral with an external source term:
$$
\mathrm{W}[v_{\mathrm{class}}]=\int\,[\mathscr{D}x(\tau)]
\exp\biggl\{\frac{i}{\hbar}\int\limits_{t_0}^{t_1}d\tau\bigl\{\tfrac{1}{2}\,\bigl(\dot{x}(\tau)\bigr)^2-\kappa\,x(\tau)\,v_{\mathrm{class}}(\tau)\bigr\}\biggr\}\,.
$$
When one performs a shift transformation:\footnote{There is an apparent ambiguity in the initial condition for the shift
function second-order differential equation $\ddot{c}(\tau)=-\kappa\,v_{\mathrm{class}}(\tau)$, but one immediately
verifies that it does not affect the value of the considered functional integral $\mathrm{W}[v_{\mathrm{class}}]$.}
$$
y(\tau)\mapsto x(\tau)=y(\tau)+c(\tau)\,,\ \ \mbox{where}\ \ \ddot{c}(\tau)=-\kappa\,v_{\mathrm{class}}(\tau)\,,
$$
then the functional measure is unchanged and the external source term gets off. Therefore we can write
$$
\mathrm{W}[v_{\mathrm{class}}]=\exp\biggl\{\frac{i}{\hbar}\int\limits_{t_0}^{t_1}d\tau\Bigl\{\tfrac{1}{2}\,\bigl(\dot{c}(\tau)\bigr)^2-\kappa\,c(\tau)\,v_{\mathrm{class}}(\tau)\Bigr\} + \bigl\{y_1\dot{c}(t_1)-y_0\dot{c}(t_0)\bigr\}\biggr\}\,
\int\,[\mathscr{D}y(\tau)]
\exp\biggl\{\frac{i}{\hbar}\int\limits_{t_0}^{t_1}d\tau\bigl\{\tfrac{1}{2}\,\bigl(\dot{y}(\tau)\bigr)^2\biggr\}\,.
$$
The last functional term is the Schr\"odinger propagator in the path integral form, which describes the motion
of a free quantal particle in between the endpoints $(q_0-c(t_0),t_0)$ and $(q_1-c(t_1),t_1)$. When putting together
all fragments that enter the formula (\ref{QMII}) and taking into account the required normalization conditions, we
arrive to the following probability amplitude:
\begin{equation}\label{friction}
\mathbf{A}(q_0,t_0;q_1,t_1)=\frac{1}{\sqrt{\smash[b]{2\pi i\hbar\Lambda}}}\exp\Bigl\{{\frac{i}{2\hbar\Lambda}(q_1-q_0)^2}\Bigr\}\,,
\ \ \mbox{where}\ \
\Lambda=\frac{2}{\kappa}\,\tanh\Bigl\{\frac{\kappa}{2}\,(t_1-t_0)\Bigr\}\,.
\end{equation}
A short inspection of (\ref{friction}) discloses that in the frictionless limit $\Lambda=(t_1-t_0)$ and the amplitude
$\mathbf{A}(q_0,t_0;q_1,t_1)$ coincides with the ordinary quantum propagator for a free particle. Moreover, it is
now explicitly acknowledged that the standard composition law for the transition amplitudes does not
operate.

\begin{figure}[tbh]
\begin{center}
\psfrag{x}{$t_1-t_0$}
\psfrag{y}{$\Lambda$}
\epsfxsize=10cm
\epsfbox{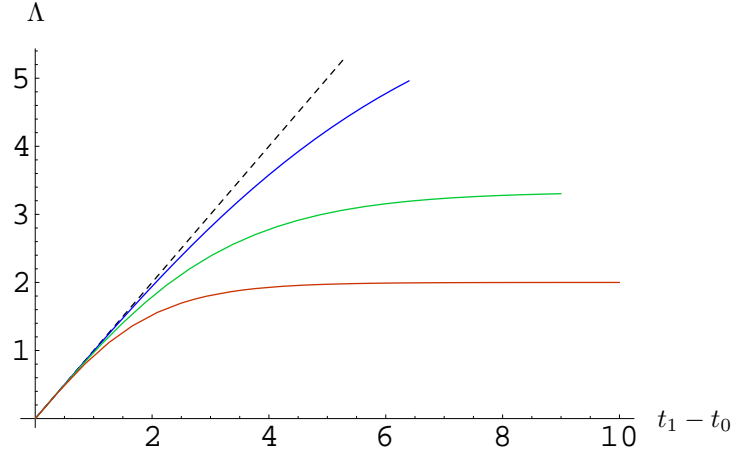}
\hspace{-3cm}
\caption{Dependence of ${\Lambda}$ on $(t_1-t_0)$ for various values of friction (red line $\kappa=1$,
green line $\kappa=0.6$, blue line $\kappa=0.3$ and dashed line $\kappa=0$).}\label{denis}
\end{center}
\end{figure}

Let us conclude this subparagraph by performing an analysis of the time evolution in terms of the transition probability
amplitude (\ref{friction}). From the point of view of quantum mechanics, the best fit of a unit mass particle with the
classical initial condition $(q_0=0, v_0=v_0, t_0=0)$ represents the gaussian wave-packet
$$
\Psi(x)\propto\exp\Bigl\{-\frac{x^2}{2\xi^2}+\frac{i}{\hbar}\,x\,v_0\Bigr\}
$$
with some initial width $\xi$. At a later time $t$, the system under the consideration will be characterized by the
convoluted wave-packet distribution
$$
\Psi(x,t)\propto\int\limits_{-\infty}^{+\infty}\,dq\,\Psi(q)\,\mathbf{A}(q,0;x,t)\,.
$$
It is clear from the properties of the convolution that the evolved wave-function $\Psi(x,t)$ preserves the gaussian
shape (modulo phase factors). Its main characteristics, the mean value of the position $\langle x\rangle$ and the
actual width of the wave-packet $\Xi^2$, are varying with the time according to
$$
\langle x\rangle=v_0\,\Lambda\ \ \ \ \ \mbox{and}\ \ \ \ \ \Xi^2=\xi^2+\frac{\hbar^2}{\xi^2}\,\Lambda^{2}\,.
$$
The velocity of the center of the wave-packed $\tfrac{d}{dt}\langle x\rangle=\tfrac{4\,v_0\,\mathrm{e}^{-\kappa t}}{(1+\mathrm{e}^{-\kappa t})^2}\propto v_0\,\mathrm{e}^{-\kappa t}$,
i.e. it decreases for $t\gg 1$ ex\-po\-nen\-tial\-ly, as one would predict on classical intuition.

\subsection{Damped harmonic oscillator}

The probability amplitude for a damped harmonic oscillator with a unit mass asks for a solution of the Newton's equation
$\ddot{x}=-\omega^2\,x-\kappa\,\dot{x}$:
$$
x_{\mathrm{class}}(\tau)=\mathrm{e}^{-\tfrac{\kappa}{2}\tau}\bigl\{A\,\mathrm{e}^{i\Theta\tau}+B\,\mathrm{e}^{-i\Theta\tau}\bigr\}\,,
\ \ \mbox{where the new frequency}\ \Theta=\sqrt{\omega^2-\tfrac{\kappa^2}{4}}\,.
$$
The dependence of $A$ and $B$ on the initial and final events $(q_0,t_0)$ and $(q_1,t_1)$ is a bit too awkward to be
presented here explicitly.
When substituting the general $\mathbb{U}(x)$ in (\ref{QMII}) by the oscillator potential
$\tfrac{1}{2}\,\omega^2\,x^2$ and after performing a similar shift business as in the above example, we arrive to:
$$
\mathbf{A}(q_0,t_0;q_1,t_1)\propto\exp\Bigl\{\frac{i}{\hbar}\,\int\limits_{t_0}^{t_1}\,d\tau\,\bigl\{\tfrac{1}{2}\,\dot{x}^2_{\mathrm{class}}-\tfrac{1}{2}\,\omega^2\,x^2_{\mathrm{class}}\bigr\}\Bigr\}
$$
Integrating the exponent and imposing the normalization conditions provide us with the propagator:
\begin{equation}\label{LHO}
\mathbf{A}(q_0,t_0;q_1,t_1)=\mathbb{K}(t_1-t_0)\,\exp\Bigl\{\frac{i}{\hbar}\,\dfrac{\Theta}{2\sin\{\Theta(t_1-t_0)\}}
\bigl[(q_1^2+q_0^2)\cos\{\Theta(t_1-t_0)\}-2(q_1q_0)\coth\{\tfrac{\kappa}{2}(t_1-t_0)\}\bigr]\Bigr\}
\end{equation}
where the normalization factor
\begin{equation}
\mathbb{K}(t_1-t_0)=\sqrt{\dfrac{\Theta\,\coth\{\tfrac{\kappa}{2}(t_1-t_0)\}}{2\pi i\hbar\,\sin\{\Theta(t_1-t_0)\}}}\,.
\end{equation}
It is clear that
\begin{itemize}
\item[$\circ$] taking the limit $\omega\rightarrow 0$; the above amplitude
reproduces the free particle result (\ref{friction}),
\item[$\circ$] in the limit $\kappa\rightarrow 0$; the Schr\"odinger propagator for a harmonic oscillator with the
frequency $\omega$ is recovered.
\end{itemize}
Let us mention that our formula (\ref{LHO}) disagrees with a result of the paper.\cite{jannussis} There is presented a
different approach to quantum mechanics in the presence of friction (time dependent formalism). Main disagreement
consists in the fact that their propagator is not invariant with respect to time translations, however, the classical
equations of motion, as well as our formula, remain invariant.

\section*{Acknowledgement}

Many thanks go to Pavel B\'{o}na, Mari\'{a}n Fecko, Tam\' as F\" ul\" op, Peter Pre\v{s}najder, Pavol \v Severa and
Vladim\' ir Balek for their interest, criticism, fruitful discussions and many useful comments. This research was
supported in part by Comenius University Grant UK/359/2006, VEGA Grant 1/3042/06 and ESF project JPD3 BA-2005/1-034.
Special thanks go to my wife Tulka for her sympathy and encouragement.\\

\centerline{$\mathscr{A.}$\ \ \ \ \ \ \ $\mathscr{M.}$\ \ \ \ \ \ \ $\mathscr{D.}$\ \ \ \ \ \ \ $\mathscr{G.}$}


\end{document}